\documentstyle[preprint,revtex]{aps}
\newcommand{\doublespace}{\renewcommand{\baselinestretch}{1.75}
   \Large\normalsize}
\begin{document}
\begin{title}
\begin{center}Superfluid characteristics of the attractive Hubbard model
for various lattice structures
\end{center}
\end{title}
\author{W. R. Czart, T. Kostyrko \& S. Robaszkiewicz}
\begin{instit}
Institute of Physics, A.Mickiewicz University,
Umultowska 85, 61-614 Pozna\'{n}, POLAND.\\

\end{instit}

\begin{abstract}
We study the basic thermodynamic and electromagnetic properties of the
superconductor described by the negative--$U$ Hubbard model (
gap parameter $\Delta$, critical temperature $T_{\mbox{\tiny C}}$, London 
penetration depth $\lambda$,
thermodynamic critical field $H_{\mbox{\tiny C}}$ and Ginzburg--Landau 
correlation length
$\xi_{\mbox{\tiny G--L}}$).

The calculations are performed for square (SQ, d=2), simple cubic (SC, d=3)
and face--centered cubic (FCC, d=3) lattices. We
analyze the results as a function of electron density $n$ and interaction $U$
using the Hartree--Fock approximation in the weak--to--intermediate $U$ case
combined with the conclusions from the perturbation theory valid in the
strong coupling limit. From the equality: $\lambda/\xi_{\mbox{\tiny G--L}}
\!=\!1/\sqrt{2}$ we
find the boundaries between local and nonlocal electronic behaviour of the
$U<0$ Hubbard model in the $U-n$ parameter space.
Using the calculated values of $T_{\mbox{\tiny C}}$ and $\lambda$ we
compare the `universal plot' of $T_{\mbox{\tiny C}}/T_{\mbox{\tiny
C}}^{\mbox{\scriptsize m}}$ vs.
$[\lambda(0)/\lambda^{\mbox{\scriptsize m}}(0)]^2$ ( where:
$T_{\mbox{\tiny C}}^{\mbox{\scriptsize m}}$ denotes maximum $T_{\mbox{\tiny
C}}$ as a function of $n$ and
$\lambda^{\mbox{\scriptsize m}}(0)$ corresponds to
$T_{\mbox{\tiny C}}^{\mbox{\scriptsize m}}$)
with the recent experimental data for various families of nonconventional
superconductors
and conclude that the best agreement can be
obtained for intermediate values of the local attraction. \\[1em]
\end{abstract}

\newpage
\doublespace
\section{Introduction}
Among the recently studied superconducting compounds there are
 several groups of
materials (copper oxides, fullerides, Chevrel phases, barium bismuthates ...)
which share some nonconventional features like high $T_{\mbox{\tiny C}}$
with relatively low carrier concentration, short coherence length together with
extremely small $\xi_{\mbox{\tiny G--L}}/\lambda$ ratio and universal
dependence of $T_{\mbox{\tiny C}}/T_{\mbox{\tiny C}}^{\mbox{\scriptsize m}}$
vs. $[\lambda(0)/\lambda^{\mbox{\scriptsize m}}(0)]^2$ 
\cite{Uemura-93,Schneider-92}.
Above features suggest that a short range almost nonretarded attraction
is responsible for pairing in these systems
\cite{Micnas-90}. One of the simplest effective models to
describe this situation is the negative--$U$ Hubbard hamiltonian. This model
constitutes a common basis for the description of the superconductors with weak
local electron pairing, being in many ways similar to the classical BCS
systems, and superconductors with strong attraction where the local pairs
conform to weakly charged hard core bosons \cite{Micnas-90}.
It has been considered as an effective model of superconductivity in the
family of cuprates 
\cite{Schneider-93,Wilson-87,Mott-93}, the barium bismuthates
\cite{Taraphder-95,Varma-89}, the fullerides
\cite{Sarker-94,Zhang-91,Chakravarty-91,Chakravarty-1991,Wilson-91}
and the Chevrel phases  \cite{Micnas-90}.

In our paper we present some new results of an analysis of the several basic
thermodynamic and electromagnetic properties of this model for
various types of lattice structure. We have
concentrated on the case of (i) square (SQ, $d$$=$$2$), (ii) simple cubic (SC,
$d$$=$$3$) and (iii) face--centered cubic (FCC, $d$$=$$3$) lattices, i.e.
structures 
representing (i) the family of cuprates, (ii) the Chevrel phases and barium
bismuthates and (iii) the fullerides, respectively.
For comparison we also present the results obtained for a model rectangular
density of states ${\cal D}(\varepsilon)$.
To calculate quantities of interest we use
the (broken symmetry) Hartree--Fock approximation (HFA) which is known to give
credible results 
at $T\!=\!0$~K as concerns energy of the ground state, the energy gap,
the chemical potential as well as the collective excitations in the whole
interaction range, 
interpolating smoothly between the weak and strong--$U$ limit, where it
matches the results of the perturbation theory developed from the
zero--bandwidth limit \cite{Kostyrko-92}.
 On the other hand the HFA leads to qualitatively
erroneous results for the the thermodynamic critical field $H_{\mbox{\tiny C}}$
 and
Ginzburg--Landau correlation length $\xi_{\mbox{\tiny G--L}}$ in the
strong attraction limit. This failure is due to the fact that the HFA greatly
overestimates the energy of the {\em normal} state which is used in a standard
calculation of $H_{\mbox{\tiny C}}$ and next $\xi_{\mbox{\tiny G--L}}$. In
order to 
discuss the behaviour of $H_{\mbox{\tiny C}}$, $\xi_{\mbox{\tiny G--L}}$ and
Ginzburg 
ratio $\kappa \!=\!\frac{\lambda}{\xi}$ in the
whole interaction range we have completed the results of the HFA calculation
with the ones of the perturbation theory. We will also point out a possible
relevance of our results to the recent experimental data for the short--coherence length
superconductors.

\section{General formulation}
We consider the negative--$U$ Hubbard Hamiltonian in the hypercubic lattice
\begin{equation}\label{Hubbard}
 H \!=\! t \sum_{ \vec{m}\vec{\delta}\sigma } \left( 
              \mbox{\rm e}^{i\Phi_{mm+\delta}} 
c^{\mbox{\scriptsize +}}_{\vec{m}\sigma } 
c^{\phantom{ }}_{ \vec{m}+{\vec{\delta}}\sigma }\! +\!
\mbox{ H.c.} \right)
- |U| \sum_{ \vec{m} } n_{\vec{m}\uparrow} n_{\vec{m}\downarrow}\,\,\,\,\,\, ,
\end{equation}
where $t$ is the hopping integral, $\vec{\delta}$ is a unit lattice vector
and the exponential (Peierls) factors in (\ref{Hubbard}) account for the
coupling of 
electrons to the magnetic field via its vector potential
$\vec{A}(\vec{r})$.  
$\Phi_{mm+\delta}\!=\!-\frac{e}{\hbar c}\vec{A}(\vec{m}) \vec{\delta}$,
and $e$ is the electron charge.
 
For weak vector potential the expectation value
of the Fourier transform of the total current operator in $\alpha$ direction
$(\alpha\!=\!x,y,z)$ can be obtained from
the linear response theory \cite{Fetter-71,Scalapino-93} as
\begin{equation}\label{response}
J_{\alpha}(\vec{q}\,,\omega) \!=\!N \frac{c}{4\pi}
\sum_{\beta} \left[
\delta_{\alpha\beta} K^{\mbox{\scriptsize \it dia}} +
K^{\mbox{\scriptsize \it para}}_{\alpha\beta}(\vec{q}\,,\omega) 
            \right]
A_{\beta}(\vec{q}\,,\omega)\,\,\,\,\, .
\end{equation}
Here we have separated the total response kernel $K_{\alpha
\beta}(\vec{q},\omega)$ into a
diamagnetic part and a paramagnetic part. The diamagnetic contribution
evaluated within HFA is 
\begin{eqnarray}\label{K_dia}
K^{\mbox{\scriptsize \it dia}} &\!=\!&
  \frac{8\pi e^2 |t|}{\hbar^2 c^2 a_{\mbox{\tiny $\bot$}}} \frac{1}{N}
 \sum_{\vec{k}\sigma}\! <c_{\vec{k}\sigma}^{+}c_{\vec{k}\sigma}> 
 \cos k_\alpha\!=\!\nonumber
\\
&&\!=\!\frac{8\pi e^2 |t|}{\hbar^2 c^2 a_{\mbox{\tiny $\bot$}}} \frac{1}{N} 
 \sum_{\vec{k}}\!
\left[\! 1\! -\!
\frac{ \lambda_{\vec{k}} }{ E_{\vec{k}} }
\tanh(\beta E_{\vec{k}}/2 )\!
\right]\cos k_\alpha \,\,\,\,\, ,
\end{eqnarray}
where $\lambda_{\vec{k}}\!=\!\epsilon_{\vec{k}}-\bar{\mu}$,
$\epsilon_{\vec{k}}\!=\!t \sum_{\vec{\delta}}\mbox{\rm e}^{i\vec{k}\vec{\delta}}$
and
$E_{\vec{k}}\!=\!\sqrt{\lambda_{\vec{k}}^2+\Delta^2}$
is the quasiparticle energy with the superconducting gap parameter \\
$\Delta\!=\!|U|x_0\!=\!\frac{|U|}{N}\sum_{i}~<c_{i\uparrow}^{+}c_{i\downarrow}^{+}>\!=\!~
\frac{|U|}{N}\sum_{\vec{k}}~<c_{\vec{k}\uparrow}^{+}c_{-\vec{k}\downarrow}^{+}>$.
The chemical potential is $\mu\!=\!\bar{\mu}-|U|n/2$, $N$ and $n\!=\!\frac{N_e}{N}$
denote the number of lattice sites and the number of electrons per site,
respectively and $0<n<2$.
The values of $\Delta$ and $\mu$ are obtained as a solution of the system of
the selfconsistent equations \cite{Robaszkiewicz-81}:
\begin{eqnarray}
\label{Delta}
\Delta &\!=\!& \frac{|U|\Delta}{2N} \sum_{\vec{k}} 
\tanh\left( \beta E_{\vec{k}}/2 \right)/E_{\vec{k}}  \,\,\,\,\, , \\
\label{mu}
n &\!=\!& \frac{1}{N} \sum_{\vec{k}}
\left[ 1 - \lambda_{\vec{k}} 
\tanh\left( \beta E_{\vec{k}}/2 \right)/E_{\vec{k}} \right]  \,\,\,\, ,
\end{eqnarray}
whereas the free energy per site of the superconducting (SS) phase and the
normal (NO) phase is given by:
\begin{eqnarray}\label{FreeSS}
\frac{F^{\mbox{\tiny SS}}}{N}\!=\!\bar{\mu}_{\mbox{\tiny SS}}
(n-1)+\frac{\Delta^2}{|U|}-\frac{|U|}{4}n^2-  
\frac{2}{\beta N}\sum_{\vec{k}}\ln 2\cosh \frac{\beta E_{\vec{k}}}{2}
\,\,\,\,\, ,
 \\
\label{FreeNO}
\frac{F^{\mbox{\tiny NO}}}{N}\!=\!\bar{\mu}_{\mbox{\tiny NO}}
(n-1)-\frac{|U|}{4}n^2- 
\frac{2}{\beta N}\sum_{\vec{k}}\ln 2\cosh \frac{\beta \lambda_{\vec{k}}}{2}
\,\,\,\,\, ,
\end{eqnarray}
where $\bar{\mu}_{\mbox{\tiny NO}}$ is determined from:
\begin{equation}
n\!=\!\frac{1}{N}\sum_{\vec{k}}\left(1-\tanh\frac{\beta
\lambda_{\vec{k}}}{2}\right) \,\,\,\,\, .
\end{equation}


The expression for the paramagnetic part of the kernel reads
\begin{eqnarray}\label{Kernel}
& &K_{\alpha \alpha^{''}}^{\mbox{\scriptsize \it para}}
(\vec{q},\omega)\!=\!-\frac{1}{N}\frac{16 \pi 
e^{2}t^{2}}{\hbar^2 c^2}
\sum_{\vec{k}} \sin\vec{k}_{\alpha}\sin\vec{k}_{\alpha^{''}} \left[ 
\phantom{\left(\frac{\bar{\epsilon}_{\vec{k}-\vec{q}}\Delta^{2}}
{E_{\vec{k}-\vec{q}}}\right) }
\right.\nonumber  \\
& &\left(1+\frac{\lambda_{\vec{k}}
\lambda_{\vec{k}-\vec{q}}+\Delta^{2}}
{E_{\vec{k}}E_{\vec{k}-\vec{q}}}\right)
\left[n_{F}(E_{\vec{k}})
-n_{F}(E_{\vec{k}-\vec{q}})\right]\times\nonumber 
                \\
& &\times\left(\frac{1}{\omega-(E_{\vec{k}-\vec{q}}
-E_{\vec{k}})+i\varepsilon}-  
\frac{1}{\omega+E_{\vec{k}-\vec{q}}
-E_{\vec{k}}+i\varepsilon}  \right)+\nonumber
                           \\
& &+\left(1-\frac{\lambda_{\vec{k}}
\lambda_{\vec{k}-\vec{q}}+\Delta^{2}}
{E_{\vec{k}}E_{\vec{k}-\vec{q}}}\right)
\left[1-n_{F}(E_{\vec{k}})
-n_{F}(E_{\vec{k}-\vec{q}})\right]\times \nonumber
                        \\
& &
\left.\times \left(\frac{1}{\omega-
(E_{\vec{k}-\vec{q}}+E_{\vec{k}})+i\varepsilon}-
\frac{1}{\omega+E_{\vec{k}}
+E_{\vec{k}-\vec{q}}+i\varepsilon} \right) 
\right]\,\,\,\,\, .
\end{eqnarray}
In the London superconductors the magnetic field penetration depth
$\lambda(T)$ is determined by the sum of diamagnetic and paramagnetic
part of the total kernel in the static limit 
$\lambda(T)\!=\!\lim_{q_y \rightarrow 0}
[-K^{\mbox{\scriptsize \it dia}}-K_{xx}^{\mbox{\scriptsize \it para}}
(q_x\!=\!0,q_y,q_z\!=\!0,\omega\!=\!0)]^{-1/2}$ \cite{Fetter-71,Scalapino-93}.

At $T\!=\!0$~K the paramagnetic part of the kernel becomes important in determining
$\lambda$ when the correlation length becomes greater than the penetration
depth and we deal in this case with nonlocal (Pippard) superconductor.
This situation is common in many low--$T_{\mbox{\tiny C}}$ systems.
The short--coherence length materials, including the high $T_{\mbox{\tiny C}}$
superconductors represent the opposite, i.e. the London limit. In the
latter 
case the ground state penetration depth is determined entirely by the
$\vec{q} \rightarrow 0$ limit of the kernel where the paramagnetic part of
the kernel vanishes and $\lambda$ is given by:

\begin{equation}\label{lambda}
\lambda \!=\! \frac{1}{\sqrt{- K^{\mbox{\scriptsize \it dia}} }}\,\,\,\,\, .
\end{equation}

In our calculation of $\lambda$ we restrict ourselves to the London
limit having in mind the properties of the systems of interest and finaly
determine the area in the $U$--$n$ parameters space where the local
approximation may be valid.
The value of the penetration depth calculated in this way is
qualitatively good both in the weak and strong--$U$ limits, in the latter case
approaching the results of the perturbation theory, as will be shown below.

Using the value of the penetration depth and the difference of the
free energy between normal and superconducting
phase one is able to determine the thermodynamic
critical field $H_{\mbox{\tiny C}}$ and the Ginzburg--Landau
correlation length $\xi_{\mbox{\tiny G--L}}$ as
\begin{eqnarray}
\label{Hc}
\frac{H_{\mbox{\tiny C}}^2(T)}{8\pi}&\!=\!&
\frac{F^{\mbox{\tiny NO}}(T)- F^{\mbox{\tiny SS}}(T)}
{Na_{\mbox{\tiny $\bot$}}a_{\mbox{\tiny $\|$}}^2} \,\,\,\,\,  ,\\
\label{xi}
\xi_{\mbox{\tiny G--L}}&\!=\!&\frac{\Phi_0}{2\pi\sqrt{2} \lambda 
H_{\mbox{\tiny C}}}
\,\,\,\,\,   ,
\end{eqnarray}
where $\Phi_0\!=\!\frac{hc}{2e}$,
and to obtain the estimations for the critical fields
$H_{\mbox{\tiny c1}}\simeq
\frac{\mbox{\scriptsize \rm ln}\kappa}\kappa
H_{\mbox{\tiny C}}\sim \frac{\mbox{\scriptsize \rm ln}\kappa}{\lambda^2}$ and 
$H_{\mbox{\tiny c2}}\simeq
\frac{\Phi_0}{2\pi\xi^2_{\mbox{\tiny G--L}}}$
, where $\kappa \!=\!\frac{\lambda}{\xi_{\mbox{\tiny G--L}}}$. At $T\!=\!0,
\frac{F(0)}{N}\!=\!E_0\!=\!\frac{1}{N}\langle H 
\rangle_{(T\!=\!0)}$. 
Within HFA the expressions determining $E_{0}^{\mbox{\tiny NO}}$ and
$E_{0}^{\mbox{\tiny SS}}$ are given by:

\begin{eqnarray}
\label{Eno}
E^{\mbox{\tiny NO}}_{0}\!=\!\frac{F^{\mbox{\tiny NO}}(0)}{N}
\!=\!E^{\mbox{\tiny K}}_{\mbox{\tiny NO}}(0)-\frac{1}{4}|U|n^2 \,\,\,\,\,  ,
\\
\label{Ess}
E^{\mbox{\tiny SS}}_{0}\!=\!\frac{F^{\mbox{\tiny SS}}(0)}{N}
\!=\!E^{\mbox{\tiny K}}_{\mbox{\tiny SS}}(0)-\frac{\Delta^2}{|U|}-\frac{1}{4}|U|n^2
\,\,\,\,\,   ,
\end{eqnarray}
where 
\begin{equation}
E^{\mbox{\tiny K}}_{\mbox{\tiny NO(SS)}}\!=\!\frac{1}{N}\sum_{\vec{k}\sigma}
\epsilon_{\vec{k}}\langle c^+_{\vec{k}\sigma}c_{\vec{k}\sigma}\rangle
_{\mbox{\tiny NO(SS)}}\,\,\,\,\, ,
\end{equation}
is the average value of the kinetic energy term in the NO(SS) phase.

The HFA calculation of the energy of the ordered state at $T\!=\!0$~K are reliable
for any $U$ \cite{Micnas-90}  and $E^{\mbox{\tiny
SS}}_{\mbox{\scriptsize 0}}$ reduces 
correctly to the exact value: $Un/2$ in 
the zero bandwidth 
limit where the electrons form a system of on--site pairs and singly occupancy
of sites is prohibited. On the other hand the HFA energy of the normal phase in
this limit is equal to $Un/4$ instead of $Un/2$ thereby producing the incorrect
energy difference of $Un/4$.
In order to remedy this inconsistency, at least in part, we resort to the
results of the perturbation theory in the calculation of
$E_0^{\mbox{\tiny NO}}$ 
for $|U|\gg t$. The hamiltonian (\ref{Hubbard}) reduces in this case to the 
pseudospin--$1/2$ Heisenberg model working in the subspace of states with no
singly occupied sites \cite{Micnas-90,Robaszkiewicz-94,Micnas-95}:
\newpage
\begin{eqnarray}\label{model-eff}
\mbox{\~{H}} &\!=\!&
-\frac{|U|}{2}\sum_{\vec{j}} \left( 2\rho^{z}_{\vec{j}} + 1 \right) 
-\frac{1}{2} \sum_{\vec{j},\vec{\delta}}
J_{\vec{\delta}} \left(
\rho^{+}_{\vec{j}}\rho^{-}_{\vec{j}+\vec{\delta}} {\rm e}^{ -2i\frac{e}
{\hbar c} \vec{A}(\vec{j})\vec{\delta}} + {\rm H.c.} \right)
\nonumber  \\
& & + \sum_{\vec{j},\vec{\delta}} J_{\vec{\delta}}
\rho^{z}_{\vec{j}} \rho^{z}_{\vec{j}+\vec{\delta}}
-\frac{N}{4} ZJ \,\,\,\,\, ,
\end{eqnarray}
where $2\rho_i^z\!=\!(n_{i\uparrow}+n_{i\downarrow}-1)$,
$\rho_i^{+}\!=\!c^{+}_{i\uparrow}c^{+}_{i\downarrow}$,
$J_{\vec{\delta}}\!=\!2t^2/|U|$ and $Z$ denotes a number 
of nearest neighbours.
The hard--core boson operators: $\rho^{\pm}_{\vec{j}},\;\rho^{z}_{\vec{j}}$
satisfy the commutation rules of the $s\!=\!\frac{1}{2}$ operators. The electron
number condition is:
\begin{equation}\label{U>>t-number_cond}
n \!=\! \frac{1}{N}\sum_{\vec{j}} \langle 2\rho^{z}_{\vec{j}} + 1 \rangle
\,\,\,\,\, ,
\end{equation}
For $x_{0}\!=\!\langle \rho^+ \rangle\!=\!
\frac{1}{N}\sum_i\langle \rho_{i}^{+} \rangle \neq 0$ the
hamiltonian (\ref{model-eff}) 
describes the superconducting state and the MFA result for the
$E_{0}^{\mbox{\tiny SS}}$ is equal to the corresponding HFA one up
to terms of order $t^2/|U|$ inclusive (see Sec.3). The MFA calculation of the
ground state 
energy for the normal state, $x_{0} \!=\! 0$, gives:
\begin{equation}\label{EN-strongU}
\bar{E}_0^{\mbox{\tiny NO}} \!=\!\frac{\langle \mbox{\~{
H}}\rangle}{N}\!=\! -\frac{1}{2}|U|n 
+ \frac{1}{4} ZJn(n-2) \,\,\,\,\, ,
\end{equation}
Using this expresion in the analysis one should keep in mind that
the MFA neglects the intersite correlation of
fluctuation of the pseudospin operators what can lead to the error of order of
$J\!=\!2t^2/|U|$.
\section{Analytical results for weak and strong $U$ cases at $T\!=\!0$~K}
The system of the selfconsistent equations (\ref{Delta},\ref{mu}) can be
approximately solved in the limiting cases of the weak and strong attraction
limits and here we present these results:
\begin{itemize}
\item[A)] The weak--$U$ case\\
Provided that the chemical potential is not located too close to van Hove
singularity of the density of states (DOS) function ${\cal D}(\varepsilon)$,
the gap parameter at $T\!=\!0$ can be obtained from equation (\ref{Delta}) as:
\begin{equation}\label{Delta-weakU}
\Delta\!=\!2 \sqrt{D_{\mbox{\scriptsize +}}D_{\mbox{\scriptsize -}}}
\exp\left(\frac{I[{\cal C}^{\bar{\mu}}]} 
{2{\cal D}(\bar{\mu})}\right) \exp\left(-\frac{1}{|U|{\cal
D}(\bar{\mu})}\right) \,\,\,\,\, ,
\end{equation}
where: $D_{\mbox{\scriptsize +}}\!=\!D_{\mbox{\scriptsize 2}}-\bar{\mu}$,
$D_{\mbox{\scriptsize -}}\!=\!D_{\mbox{\scriptsize 1}}+\bar{\mu}$ with
$-D_{\mbox{\scriptsize 1}}$,$D_{\mbox{\scriptsize 2}}$
denoting lower and upper boundary of the band, respectively;
$I[{\cal C}^{\bar{\mu}}]$
is a functional defined by

\begin{equation}
I[f^{\bar{\mu}}]\!=\!\int_{-D_1}^{D_2} d\epsilon
\mbox{\rm sign}(\epsilon-\bar{\mu})f^{\bar{\mu}}(\epsilon) \,\,\,\,\, ,
\end{equation}

${\cal C}^{\bar{\mu}}$ is a function of electron density, which depends on
details of DOS and is defined by
\begin{equation}\label{Cmu}
{\cal C}^{\bar{\mu}}(\epsilon)\!=\!
\frac{{\cal D}(\epsilon)-{\cal D}(\bar{\mu})}{\epsilon-\bar{\mu}}
\,\,\,\,\, .
\end{equation}
In Eqs.(\ref{Delta-weakU}),(\ref{Cmu}) the chemical potential may
be approximated by its value in the normal state,
$\bar{\mu}_{\mbox{\tiny NO}}$ and it is determined by:
\begin{equation}\label{muN}
n-1\!=\!-\int_{-D_{\mbox{\scriptsize 1}}}^{D_{\mbox{\scriptsize 2}}}
d\epsilon  
{\cal D}(\epsilon)\mbox{sign}(\epsilon-\bar{\mu}_{\mbox{\scriptsize N}})
\,\,\,\,\, .
\end{equation}
The critical temperature in this limit is given by:
\begin{equation}\label{Tc-weakU}
k_{\mbox{\tiny B}} T_{\mbox{\tiny C}}/\Delta \approx 1/1.76
\,\,\,\,\, ,
\end{equation}
which is the well--known weak coupling BCS ratio.

Analogously, performing the weak coupling expansions of Eqs.(\ref{K_dia},
\ref{mu}, \ref{Hc})
at $T\!=\!0$ one can find the analytical expressions determining
$\lambda$, $\bar{\mu}_{\mbox{\tiny SS}}$, $H_{\mbox{\tiny C}}$,
$\xi_{\mbox{\tiny G--L}}$ and
$\kappa$.
The London penetration depth is calculated as
\begin{equation}\label{lambda_Uweak}
\frac{1}{\lambda^2(0)} \!=\!-K^{\mbox{\scriptsize \it dia}}\!=\!
-\frac{8\pi e^2}{\hbar^2 c^2 a_{\mbox{\tiny $\bot$}} }\frac{1}{Z}
E^{\mbox{\tiny K}}_{\mbox{\tiny SS}}(0) \,\,\,\,\, ,
\end{equation}
where

\begin{eqnarray}\label{Ekss_Uweak}
E^{\mbox{\tiny K}}_{\mbox{\tiny SS}}(0)-E^{\mbox{\tiny K}}_{\mbox{\tiny NO}}
(0)&\!=\!&
\frac{\Delta^2}{|U|}
-\Delta^2\left\{\frac{E^{\mbox{\tiny K}}_{\mbox{\tiny
NO}}-\bar{\mu}n}{4}\right.\times \nonumber
\\
\times&&\left[\frac{1}{(D_2-\bar{\mu})^2}+\frac{1}{(D_1+\bar{\mu})^2}
 \right]
+\frac{\bar{\mu}}{2(D_2-\bar{\mu})^2}\nonumber
\\
+&&\left.{\cal D}(\bar{\mu})+\frac{1}{2}I\left[{\cal C}^{\bar{\mu}}\right]
-I\left[{\cal F}_3^{\bar{\mu}}\right]+\bar{\mu}
I\left[{\cal G}_3^{\bar{\mu}}\right]\right\}
\,\,\,\,\,  ,
\end{eqnarray}
where
\begin{eqnarray}
{\cal G}_{n+1}^{\bar{\mu}}(\epsilon)&\!=\!&
\frac{{\cal G}^{\bar{\mu}}_n(\epsilon)-{\cal
G}^{\bar{\mu}}_n(\bar{\mu})}{\epsilon - \bar{\mu}}\,\,\,\,\,\, ,
\,\,\,\,\,\,\,\,\,\,\,\,\,\,\,\,\,\, 
{\cal G}_0(\epsilon)\!=\!\int_{-D_1}^{\epsilon} d\epsilon'{\cal
D}(\epsilon')\epsilon' \nonumber
\\
{\cal F}_{n+1}^{\bar{\mu}}(\epsilon)&\!=\!&
\frac{{\cal F}^{\bar{\mu}}_n(\epsilon)-{\cal
F}^{\bar{\mu}}_n(\bar{\mu})}{\epsilon - \bar{\mu}}\,\,\,\,\,\, ,
\,\,\,\,\,\,\,\,\,\,\,\,\,\,\,\,\,\,
{\cal F}_0(\epsilon)\!=\!\int_{-D_1}^{\epsilon} d\epsilon' {\cal
D}(\epsilon')
\end{eqnarray}
and $\bar{\mu}\!=\!\bar{\mu}_{\mbox{\tiny NO}}$. Obviously, for $|U|/B
\rightarrow 0 \;\;
E^{\mbox{\tiny K}}_{\mbox{\tiny SS}}(0)$
approaches the value of the band energy in the normal state.

In a case when we can neglect variation of ${\cal D}(\epsilon)$ 
and $D_1\!=\!D_2\!=\!D$ Eq.(\ref{Ekss_Uweak}) reduces to 

%
\begin{eqnarray}\label{Ekss_UweakD}
E^{\mbox{\tiny K}}_{\mbox{\tiny SS}}(0)&\!=\!&
-{\cal D}(\bar{\mu}_{\mbox{\tiny
NO}})\left[D^2-\bar{\mu}_{\mbox{\tiny
NO}}^2+\frac{\Delta^2}{2}
-\frac{\Delta^2}{|U|{\cal D}(\bar{\mu}_{\mbox{\tiny NO}})}\right]\,\,\,\,\, ,
\end{eqnarray}
whereas $\bar{\mu}_{\mbox{\tiny SS}}$ and $H_{\mbox{\tiny C}}(0)$ are given by

\begin{equation}\label{muss_Uweak}
\bar{\mu}_{\mbox{\tiny SS}}^2 \!=\! \bar{\mu}_{\mbox{\tiny NO}}^2\left(
1+\frac{\Delta^2}{D^2-\bar{\mu}^2_{\mbox{\tiny NO}}}\right) \,\,\,\,\, ,
\end{equation}
where $\bar{\mu}_{\mbox{\tiny NO}}$ is given by Eq.(\ref{muN}), and

\begin{equation}\label{Hc_Uweak}
\frac{H_{\mbox{\tiny C}}^2(0)}{8\pi}
\!=\!\frac{E_0^{\mbox{\tiny NO}}-
E_0^{\mbox{\tiny SS}}}{a_{\mbox{\tiny $\bot$}}a^{2}_{\mbox{\tiny $\|$}}}
\approx \frac{1}{2}
\frac{\Delta^2 {\cal D}(\bar{\mu}) }{a_{\mbox{\tiny $\bot$}}
a^{2}_{\mbox{\tiny $\|$}} } \,\,\,\,\, . 
\end{equation}
The weak coupling formulas for $\xi_{\mbox{\tiny G--L}}$ and
$\kappa \!=\!\lambda/\xi_{\mbox{\tiny G--L}}$ follow directly from 
Eqs.(\ref{xi}), (\ref{lambda_Uweak}), (\ref{Hc_Uweak}).
In the case of ${\cal D}_1(\bar{\mu})\!=\!0$ and $D_1\!=\!D_2\!=\!D$ their explicit forms
are: 

\begin{equation}\label{ksi_Uweak}
\xi_{\mbox{\tiny G--L}}\!=\!\frac{a_{\mbox{\tiny $\|$}} }{2\sqrt{2Z}}
\sqrt{-\frac{D^2-\bar{\mu}_{\mbox{\tiny NO}}^2}{\Delta^2}+\frac{1}{2}-
\frac{1}{|U|{\cal D}(\bar{\mu})}}\approx
\frac{a_{\mbox{\tiny $\|$}} }{2\sqrt{2Z}}
\mbox{\rm exp}\left(\frac{1}{|U|{\cal D}(\bar{\mu})}\right) \,\,\,\,\, ,
\end{equation}

\begin{eqnarray}\label{ratio_weak}
\kappa &\!=\!&
\frac{\Phi_0 Z}{\pi\sqrt{\pi}}\frac{\sqrt{a_{\mbox{\tiny $\bot$}}}}
{a_{\mbox{\tiny $\|$}}}
\frac{\sqrt{\Delta^2 {\cal D}(\bar{\mu}_{\mbox{\tiny NO}})}}
{\left(D^2-\bar{\mu}_{\mbox{\tiny NO}}^2+\frac{\Delta^2}{2}\right){\cal
D}(\bar{\mu}_{\mbox{\tiny NO}} ) 
-\frac{\Delta^2}{|U|}}\approx\nonumber 
\\
&\approx&\frac{\Phi_0 Z}{\pi\sqrt{\pi}}\frac{\sqrt{a_{\mbox{\tiny $\bot$}}}}
{a_{\mbox{\tiny $\|$}}}
\frac{ \;\mbox{\rm exp}
\left(-1/|U|{\cal D}(\bar{\mu}_{\mbox{\tiny NO}})\right)}
{\sqrt{(D^2-\bar{\mu}_{\mbox{\tiny NO}}^2){\cal
D}(\bar{\mu}_{\mbox{\tiny NO}})}} \,\,\,\,\, .
\end{eqnarray}
\item[B)] The strong--$U$ case\\
In this case the integrands in Eqs.(\ref{Delta},\ref{mu}) can be expanded in series
with respect to $t/U$ and we have:

\begin{equation}\label{X0_Ustrong}
x_0\!=\!\frac{\Delta}{|U|} \approx
\frac{1}{2}\sqrt{1-\delta^2}\sqrt{1-4L_2\left(\frac{t}{U}\right)^2+16\delta
L_3\left(\frac{t}{|U|}\right)^3 } \,\,\,\,\, ,
\end{equation}
\begin{equation}\label{UX0_Ustrong}
\bar{\mu}_{\mbox{\tiny SS}} \!=\!
-\frac{|U|}{2}\delta\left(1+4L_{\mbox{\scriptsize 2}}(t/U)^{2}-
4L_{\mbox{\scriptsize 3}}(3\delta-1/\delta)(t/|U|)^{3}\right) \,\,\,\,\, ,
\end{equation}
where $\delta\!=\!1-n$, $\bar{\mu}_{\mbox{\tiny NO}}$ is given by (\ref{muN})
and the lattice sums $L_n$ represent the moments of DOS 
and are given by:
\begin{equation}
L_n \!=\! \sum_{\vec{\delta}_1...\vec{\delta}_n}
\delta_{\vec{\delta}_1+...\vec{\delta}_n,\vec{0}} \,\,\,\,\, .
\end{equation}
One can note that odd moments vanish for the alternating (i.e. SQ and SC)
lattices but not for the FCC ones. For alternating lattices $L_2\!=\!Z$,
 $L_3\!=\!0$
and for FCC lattice $L_2\!=\!12$ and $L_3\!=\!48$.
The corresponding formula for $\lambda$
obtained by expanding of Eq.(\ref{K_dia}) reads
\begin{equation}\label{lambda-strongU}
\frac{1}{\lambda^{2}} \!=\! \frac{8\pi e^2}{\hbar^2 c^2 a_{\mbox{\tiny $\bot$}}}
\frac{|U|}{Z}
[2(1-\delta^{2})L_{\mbox{\scriptsize 2}}(t/U)^{2}
-6\delta(1-\delta^{2})L_{\mbox{\scriptsize 3}}(t/|U|)^{3}] \,\,\,\,\, .
\end{equation}
In the structures with asymmetric DOS (like FCC) the odd terms in $t/U$ from
Eq.(\ref{lambda-strongU}) are again nonzero.
From (\ref{Eno}) and (\ref{Ess}) we obtain:
\begin{eqnarray}\label{Eno_Uweak}
E_0^{\mbox{\tiny NO}}\!=\!
-\frac{1}{N}\sum_{\vec{k}}\epsilon_{\vec{k}}\;\mbox{\rm sign}(\epsilon -
\bar{\mu}_{\mbox{\tiny NO}}) - \frac{1}{4}|U|n^2 \,\,\,\,\, ,
\end{eqnarray}
\begin{equation}\label{Ess_Ustrong}
E_0^{\mbox{\tiny SS}}
\!=\!-\frac{|U|}{2}n-n(2-n)L_2\frac{t^2}{|U|}+2n(1-n)(2-n)L_3\frac{t^3}{U^2}
\,\,\,\,\, ,
\end{equation}
\begin{equation}\label{Hc_Ustrong}
\frac{ H_{\mbox{\tiny C}}^2}{8\pi}\!=\!\frac{E_0^{\mbox{\tiny
NO}}-E_0^{\mbox{\tiny SS}}}{a_{\mbox{\tiny $\|$}}^2a_{\mbox{\tiny $\bot$}}}\!=\!
\frac{1}{a_{\mbox{\tiny $\|$}}^2
a_{\mbox{\tiny $\bot$}}}\left\{\frac{|U|}{4}n(2-n)+
n(2-n)\frac{zt^2}{|U|}\right\}+O\left(\frac{t^3}{U^2}\right) \,\,\,\,\, ,
\end{equation}
\begin{equation}\label{ksi_Ustrong}
\xi_{\mbox{\tiny
G--L}}\!=\!\frac{a_{\mbox{\tiny $\|$}}}{2}
\left|\frac{t}{U}\right|\left[\frac{1}{4}+Z\left(
\frac{t}{U}\right)^2\right]^{-1} \,\,\,\,\, ,
\end{equation}
\begin{equation}\label{ratio_Ustrong}
\kappa \!=\!
\frac{\Phi_0}{2\pi\sqrt{\pi}}\frac{\sqrt{a_{\mbox{\tiny
$\bot$}}}}{a_{\mbox{\tiny $\|$}}}
\frac{1}{\sqrt{n(2-n)}}\frac{\sqrt{|U|^3}}{4t^2}
\sqrt{1+4Z\left(\frac{t}{U}\right)^2}
\,\,\,\,\, .
\end{equation}
\end{itemize}

It is worth while to compare the strong coupling results of HFA with
the ones obtained for the effective pseudospin model Hamiltonian 
(\ref{model-eff}).
Recently, the electromagnetic properties of that model have been analysed in
\cite{Robaszkiewicz-94,Micnas-95}  and below we only quote the expressions
derived within MFA (the 
RPA treatment, taking into account quantum corrections, yields qualitatively
similar results, except the low density limit \cite{Micnas-95}):

\begin{equation}\label{MFAX0}
x_0\!=\!\frac{1}{2}\sqrt{n(2-n)} \,\,\,\,\, ,
\end{equation}
\begin{equation}\label{MFAEss}
E_0^{\mbox{\tiny SS}}\!=\!-\frac{1}{2}|U|n-\frac{1}{2} J_0 n(2-n)
\,\,\,\,\, , 
\end{equation}
\begin{equation}\label{MFAHc}
\frac{H_{\mbox{\tiny C}}^2}{8\pi}\!=\!\frac{1}{4}\frac{J_0n(2-n)}{a^3} \,\,\,\,\,
, 
\end{equation}
\begin{equation}\label{MFAlambda}
\lambda^{-2}\!=\!\frac{2\pi \bar{e}^2 }{\hbar^2 c^2
a}Jn(2-n) \,\,\,\,\,  ,
\end{equation}
where $J_0\!=\!ZJ$, $J\!=\!\frac{2t^2}{|U|}$ and $\bar{e}\!=\!2e$, $a_{\mbox{\tiny
$\bot$}}\!=\! 
a_{\mbox{\tiny $\|$}}\!=\!a$,

\begin{equation}\label{15}
\xi_{\mbox{\tiny G--L}}\!=\!\frac{a}{\sqrt{2Z}} \,\,\,\,\, ,
\end{equation}

\begin{equation}\label{151kappa}
\kappa \!=\!\frac{\lambda}{\xi_{\mbox{\tiny G--L}}}\!=\!\frac{\Phi_0
\sqrt{Z}}{2\pi\sqrt{\pi}\sqrt{aJ n(2-n)}} \,\,\,\,\, ,
\end{equation}
and $E_0^{\mbox{\tiny NO}}$ is given by Eq.(\ref{EN-strongU}).
Up to terms of order $t^2/|U|$ the HFA expressions for $x_0$
 (Eq.(\ref{X0_Ustrong})),
$E_0^{\mbox{\tiny SS}}$ (Eq.(\ref{Ess_Ustrong})) and 
$1/\lambda^2$ (Eq.(\ref{lambda-strongU})) are equal to those 
given by Eqs.(\ref{MFAX0}), (\ref{MFAEss}) and (\ref{MFAlambda}).
On the contrary the HFA results for $H_{\mbox{\tiny C}}$, $\xi_{\mbox{\tiny
G--L}}$ 
and $\kappa$ are qualitatively erroneous due to the great overestimation 
of $E_0^{\mbox{\tiny NO}}$ by HFA in this limit 
(comp. Eq.(\ref{Eno_Uweak}) and 
Eq.(\ref{EN-strongU})). Thus, to get
realistic values of $H_{\mbox{\tiny C}}$, $\xi_{\mbox{\tiny G--L}}$ and
$\kappa$ 
for arbitrary $|U|/B$ one should combine the results obtained using 
$E_0^{\mbox{\tiny NO}}$ given by Eq.(\ref{Eno}) (reliable for
$|U|/B<1$) with those obtained using $E_0^{\mbox{\tiny NO}}$
given by Eq.(\ref{EN-strongU}) (reliable for $|U|/B>1$). 
This will be done in the next section, where we will also compare some of
the predictions given above with the results of numerical solutions of 
Eqs.(\ref{K_dia})-(\ref{mu}), (\ref{Hc})-(\ref{Ess}).
\section{Results of numerical solution and discussion.}
Figs.1a and 1b show the numerical plots of $T_{\mbox{\tiny C}}(n)$,
$\Delta(n)$ and 
$\lambda^{-2}(n)$ for SC and FCC lattices, respectively (the corresponding
plots for SQ lattice are given in Fig.1 of Ref.\cite{Czart-95}), whereas Fig.1c
presents 
the concentration dependence of $\Delta$ and $\lambda^{-2}(n)$ for the model
rectangular ${\cal D}(\varepsilon)$:

\begin{equation}\label{2d}
{\cal D}(\varepsilon)\!=\!1/B \mbox{ for }
-B/2<\varepsilon<B/2\!=\!D \,\,\,\,\, , \mbox{ otherwise 0},
\end{equation}
where $B$ is the effective bandwidth. All these plots
were made for the values of $U$ from the
weak coupling regime ($|U|/B\!=\!0.1$) where the effects of the density of
states, ${\cal D}(\varepsilon)$, are most clearly seen.
In this regime the
superconducting critical
temperature $T_{\mbox{\tiny C}}$ and the $T\!\!=\!\!0$ gap parameter $\Delta$
increase rapidly with $\bar{\mu}$ 
approaching van Hove singularities in ${\cal D}(\varepsilon)$, in agreement
with a modified BCS expression Eq.(\ref{Delta-weakU}).
This is the reason why $T_{\mbox{\tiny C}}$
 for SQ and SC lattices is symmetric with respect to
$n$$=$$1$ and so strongly peaked for SQ lattice at $n$$=$$1$ 
while the position of $T_{\mbox{\tiny C}}$ 
peak in the FCC structure moves with increase of
$|U|$ from the vicinity of $n$$\approx$$2$ toward $n$$=$$1$. For the FCC
lattice the effect of density of states on the penetration depth consist
mainly in 
$U$-dependent shift of the maximum of $1/\lambda^2$ from $n\!=\!1$ towards
$n<1$ 
(Fig.1b). 
The inverse squared penetration depth, being proportional to the 
bandwidth for weak--$U$ decreases like $\sim~Zt^2/|U|$ in the strong coupling
limit. The plot of this quantity as a function of $n$ for a few values of
$|U|/B$ is shown in Fig.2 (SQ lattice).

Examples of the evolution of the thermodynamic critical field
$H_{\mbox{\tiny C}}$ and of
the correlation length $\xi_{\mbox{\tiny G--L}}$, with
$|U|$ and $n$ are shown in Fig.3,4.
Note the steep decrease of $\xi_{\mbox{\tiny G--L}}$ with $n$ in
 a relatively narrow range
of $n$ values for the FCC lattice and  substantial reduction of
$\xi_{\mbox{\tiny G--L}}$
near $n\!=\!1$ for the SQ lattice (Fig. 5). These features can be understood
on the basis of the $\Delta(n)$ behaviour presented in Figs 1,2.

The substantial variation of $\xi_{\mbox{\tiny G--L}}$
with $n$ for $|U|/B<1$ is largely
due to the strong $n$ dependence of $H_{\mbox{\tiny C}}^2$ which in
the weak coupling limit is proportional to $\Delta^2{\cal D}(\bar{\mu})$ 
(cf. Eqs.(\ref{Hc_Uweak}), (\ref{ksi_Uweak})).

With increasing $|U|$ the $n$--dependence of
the correlation length is less pronounced and in the strong coupling regime,
$|Zt/U|$$\ll$$1$, $\xi_{\mbox{\tiny G--L}}$ calculated with in HFA
goes like $\sim |Zt/U|$
being almost independent on $n$.
However, for large $|U|$ the energy of the {\em normal} phase, entering
(\ref{Hc}), (\ref{xi}) is drastically overestimated in the HFA
(comp. Sec. 3). 

To get more realistic values of $H_{\mbox{\tiny C}}$ and
$\xi_{\mbox{\tiny G--L}}$ for $|U|/B$$>$$1$ we have made an
estimation of $\xi_{\mbox{\tiny G--L}}$ and $H_{\mbox{\tiny C}}$ exploiting
perturbation 
approach in the calculation of the energy of the normal phase 
(Eq.(\ref{EN-strongU})) and the
numerical results for SQ lattice and $n\!=\!1$ are shown in Figs 3,4. 
In the $|U|/B\gg 1$ limit the analytical results for $H_{\mbox{\tiny C}}$, 
$\xi_{\mbox{\tiny G--L}}$ and $\kappa$ can be obtained in this way
by using Eqs.(\ref{EN-strongU}, \ref{Ess_Ustrong}, \ref{lambda-strongU}
and
\ref{UX0_Ustrong}):
\begin{equation}\label{Hc_Ustrong1}
\frac{ H_{\mbox{\tiny C}}^2}{8\pi}\!=\!\frac{E_0^{\mbox{\tiny
NO}}-E_0^{\mbox{\tiny SS}}}{a_{\mbox{\tiny $\|$}}^2a_{\mbox{\tiny $\bot$}}}\!=\!
\frac{1}{a_{\mbox{\tiny $\|$}}^2a_{\mbox{\tiny $\bot$}}}
\left(\frac{1}{2}n(2-n)L_2\frac{t^2}{|U|}-
2n(1-n)(2-n)L_3\frac{t^3}{U^2}\right)
 \,\,\,\,\, ,
\end{equation}
\begin{equation}\label{ksi_Ustrong1}
\xi_{\mbox{\tiny G--L}}\!=\!\frac{a_{\mbox{\tiny $\|$}}}{\sqrt{2Z}}
\sqrt{\frac{Z-3(1-n)L_3\frac{t}{|U|}}{Z-
4(1-n)L_3\frac{t}{|U|}}}
\,\,\,\,\, ,
\end{equation}
\begin{equation}\label{ratio_Ustrong1}
\kappa \!=\!
\frac{\Phi_0}{2\pi\sqrt{\pi}}\frac{Z\sqrt{a_{\mbox{\tiny
$\bot$}}}}{a_{\mbox{\tiny $\|$}}}
\frac{1}{\sqrt{n(2-n)\frac{2t^2}{|U|}}}
\frac{\sqrt{Z-4(1-n)L_3\frac{t}{|U|}}}{Z-
3(1-n)L_3\frac{t}{|U|}}
\,\,\,\,\, .
\end{equation}
They coincide (to the second order in $\frac{t}{|U|}$) with the MFA
results for the pseudospin model (\ref{model-eff}) (cf. 
Eqs.(\ref{MFAHc}), (\ref{15}), (\ref{151kappa})).

As we see from Figs 3a and 4, for $|U|/B>1$ $H_{\mbox{\tiny C}}$ decreases
with increasing 
$|U|/B$ and for
$|U|/t\gg 1$, $H_{\mbox{\tiny C}}^2\sim t^2/|U|$, 
whereas $\xi_{\mbox{\tiny G--L}}$ tends 
to a fixed value $a/\sqrt{2Z}$.

Finally let us conclude the evolution of the Ginzburg ratio $\kappa
\!=\!\lambda/\xi_{\mbox{\tiny G--L}}$ with $|U|$ and $n$. Concentration
dependence of this 
quantity is plotted in Fig 6.
Notice universal behaviours (i) $\kappa\sim n^{-1/2}$ for $n\ll 1$
(arbitrary $|U|/B$)
and (ii)
$\kappa\sim [n(2-n)]^{-1/2}$ for $|U|/t\gg 1$ (arbitrary $n$).
In the weak-to-intermediate coupling regime ($|U|/B<1$) $\kappa (n)$ is not
universal and strongly depends on the details of ${\cal D}(\epsilon)$.
From the equation $\lambda/\xi_{\mbox{\tiny G--L}}\!=\!1/\sqrt{2}$ one can
estimate the 
boundaries between local and nonlocal electromagnetic behaviour in the
considered model \cite{Fetter-71,Scalapino-93}.
In Fig. 7 we plot this boundary in the $|U|/B - n$ parameter space for fixed
value of $t$ and $a$.
In the low concentration limits ($|n-1|\sim 1$) the local behaviour extends up
to very small values of $|U|/B$.

The high-$T_{\mbox{\tiny C}}$ systems (not only the cuprates but
also the barium bismuthates and the fullerides) are extreme type II
superconductors with local electromagnetic behaviour and with very large
Ginzburg 
ratio estimated to be of order $10^2$ \cite{Micnas-90,Holczer-92}.
Thus, from the equality $\kappa \!=\!10^2$ one can estimate the values of
$|U|/B$ and 
$n$ which could be reliable for these materials.
Examples of such estimations are given in Fig. 8
and they indicate that best agreement can be obtained for intermediate values 
of the local attraction.

In order to make some more detailed comparisons with experiment we have
evaluated the plots $T_{\mbox{\tiny C}}/T_{\mbox{\tiny C}}^{\mbox{\tiny m}}$
vs.  
$(\lambda^{\mbox{\tiny m}}/\lambda)^2$, where $T_{\mbox{\tiny
C}}^{\mbox{\tiny m}}$ denotes the 
maximum critical temperature in the $0$$<$$n$$<$$2$ range and
$\lambda^{\mbox{\tiny m}}$ 
corresponds to $T_{\mbox{\tiny C}}^{\mbox{\tiny m}}$. For a given $|U|/B$ the
data 
corresponding to SQ and 
SC lattices represent the one--valued function (see Figs.9ab), whereas the
ones for the FCC 
lattice (see Fig.9c) exhibit substantial hysteresis -- the
behaviour for the FCC structure 
being due to the difference in positions of $T_{\mbox{\tiny C}}(n)$ and
$1/\lambda^2(n)$ 
maxima seen in Fig.1b. At present the available experimental data
\cite{Schneider-93,Niedermayer-93,Uemura-93,Schneider-92} correspond mainly to SQ lattices
(cuprates) and alternating cubic
lattices (Chevrel phases, doped BaBi0$_3$) and except of the overdoped cases they compare
well with the present plots for these lattices in the intermediate coupling
range ($|U|/B\leq 1$). 

The experimentally found deviations from the 
$T_{\mbox{\tiny C}}/T_{\mbox{\tiny C}}^{\mbox{\tiny m}}$
vs.  $(\lambda^{\mbox{\tiny m}}/\lambda)^2$ universal dependence in the
strongly overdoped 
regime can be explained either by the asymmetry in ${\cal D}(\varepsilon)$
(cf Fig 9c) due to further neighbour hopping in SC and SQ structures or
by the effects of 
intersite Coulomb interactions \cite{Robaszkiewicz-94}, which can be different
for various families of materials.

Very recently Locquet \cite{Locquet-95}  have
reported the
penetration depth measurements for La$_{\mbox{\scriptsize 2-x}}$
Sr$_{\mbox{\scriptsize x}}$CuO$_{\mbox{\scriptsize 4}}$ films as a
function
of doping,
extending for the first time in the range from heavily underdoped to heavily
overdoped.
As we see from Fig.10 the theoretical plots $\lambda(n)/\lambda(n\!=\!1)$ for our
simple effective model for SQ lattice fit surprisingly well these
experimental data.

\acknowledgements
This work has financial supports
from K.B.N. Poland, projects 2~P03B~104~11, 2P03B05709.\\
We are grateful to Prof. R. Micnas for very helpful discussion.
One of us (W.R.C.) wants to thank to the Center de Recherches sur les
Tr\`{e}s Basses
Temp\'{e}ratures (CRTBT) of the
Centre National de la Recherche Scientifique in Grenoble for hospitality
and to the UESCO/ICSU/TWAS for a grant Nr.  SC/RP 206.770.5 which made
possible his stay in the CRTBT during the course of this work. Thanks
are also due to the French--Polish Scientific and Technological Cooperation 
Joint Project for 1996, Proj. Nr. 6448. 

\newpage
\appendix{}
\noindent
For rectangular DOS defined by (\ref{2d}).
one can derive from Eqs.(\ref{lambda}-\ref{xi}) analytical expressions
determining $\lambda$,
$H_{\mbox{\tiny C}}$, $\xi_{\mbox{\tiny G--L}}$ and
$\kappa$
 as a function of $n$ and  
$|U|/D$ at $T\!=\!0$.\\
In particular, from Eqs.(\ref{Delta}-\ref{FreeNO}) taken at $T\!=\!0$ one obtains 
\cite{Robaszkiewicz-81}

\begin{equation}\label{DeltaDos}
\Delta \!=\!\sqrt{n(2-n)} D/\sinh \frac{2D}{|U|} \,\,\,\,\, ,
\end{equation}

\begin{equation}\label{mussDos}
\bar{\mu}_{\mbox{\tiny SS}} \!=\!(n-1)D \coth \frac{2D}{|U|} \,\,\,\,\, ,
\end{equation}

\begin{equation}\label{munoDos}
\bar{\mu}_{\mbox{\tiny NO}}\!=\!(n-1)D \,\,\,\,\, ,
\end{equation}

\begin{equation}\label{A3}
E^{\mbox{\tiny NO}}_0 \!=\! - \frac{D}{2} n(2-n) - \frac{|U|}{4}n^2
\,\,\,\,\, ,
\end{equation}

\begin{equation}\label{A4}
E^{\mbox{\tiny SS}}_0 \!=\! - \frac{D}{2} n(2-n) \coth \frac{2D}{|U|}
 - \frac{|U|}{4}n^2 \,\,\,\,\, ,
\end{equation}
whereas $1/\lambda^2(0)\!=\!-K^{\mbox{\scriptsize \it dia}}$ calculated from
Eqs.(\ref{K_dia}) and 
(\ref{Delta}), (\ref{mu}) is:

\begin{equation}\label{A5}
\frac{1}{\lambda_L^2}\!=\! \frac{4\pi e^2t}{\hbar^2 c^2 a_{\mbox{\tiny
$\bot$}}} 
  n(2-n) \left[ \coth\frac{2D}{|U|}
-\frac{2D}{|U|}\left( \sinh\frac{2D}{|U|} \right)^{-2} \right] 
\,\,\,\,\, ,
\end{equation}
Taking into account Eqs.(\ref{A3})-(\ref{A5}) in Eqs.(\ref{Hc}) and
(\ref{xi}) one finds the 
following expressions for $H_{\mbox{\tiny C}}$, $\xi_{\mbox{\tiny G--L}}$ and
$\kappa$: 

\begin{equation}\label{A6}
\frac{H_{\mbox{\tiny C}}^2}{8\pi}\!=\!
\left(\frac{\Delta_E}{a_{\mbox{\tiny $\|$}}^2(\mbox{\AA})a_{\mbox{\tiny
$\bot$}} 
(\mbox{\AA})}\right)
  \,\,\,\,\, ,
\end{equation}
\begin{equation}\label{A61}
\Delta_E\!=\!E_{0}^{\mbox{\tiny NO}}-E_{0}^{\mbox{\tiny SS}}\!=\!
\frac{Dn(2-n)}{2}\left[\coth \frac{2D}{|U|}-1\right] \,\,\,\,\, ,
\end{equation}

\begin{equation}\label{A7}
\xi_{\mbox{\tiny G--L}}\!=\!
\frac{a_{\mbox{\tiny $\|$}}}{2\sqrt{2Z}}\sqrt{\frac{\coth\frac{2D}{|U|} 
-\frac{2D}{|U|}\left(\sinh\frac{2D}{|U|}\right)^{-2}}
{\left[\coth\frac{2D}{|U|}-1\right]}} \,\,\,\,\, ,
\end{equation}
where $D\!=\!Zt$.

\begin{equation}\label{A8}
\kappa \!=\!\frac{\Phi_0\sqrt{2Z}}{\pi\sqrt{\pi}} 
\frac{\sqrt{\frac{a_{\mbox{\tiny $\bot$}}}{t}}}
{a_{\mbox{\tiny $\|$}}}
\sqrt{\frac{\coth\frac{2D}{|U|}-1}{\left[\coth\frac{2D}{|U|}
-\frac{2D}{|U|}\left(\sinh
\frac{2D}{|U|}\right)^{-2}\right]^2}}\times
\frac{1}{\sqrt{n(2-n)}} \,\,\,\,\, ,
\end{equation}
In the limits (i) $|U|/D\ll 1$ and (ii) $|U|/D \gg 1$ the above equations
take the form:\\
for (i)

\begin{equation}\label{A9}
\frac{1}{\lambda_L^2}\!=\!
\frac{4\pi e^2t}{\hbar^2 c^2 a_{\mbox{\tiny $\bot$}}} 
n(2-n)\left[1-\frac{8D}{|U|}\mbox{\rm exp}\left(-\frac{4D}{|U|}\right)\right]
\,\,\,\,\, ,
\end{equation}


\begin{equation}\label{A10}
\Delta_E\!=\!Dn(2-n)\mbox{\rm exp}\left(-\frac{4D}{|U|}\right) \,\,\,\,\, ,
\end{equation}

\begin{equation}\label{A11}
\xi_{\mbox{\tiny G--L}}\!=\!
\frac{a_{\mbox{\tiny $\|$}}}{2\sqrt{2Z}}
\mbox{\rm exp}\left(\frac{2D}{|U|}\right) \,\,\,\,\, ,
\end{equation}

\begin{equation}\label{A12}
\kappa\!=\!\frac{\Phi_0\sqrt{2Z}}{\pi\sqrt{\pi}} 
\frac{\sqrt{\frac{a_{\mbox{\tiny $\bot$}}}{t}}}
{a_{\mbox{\tiny $\|$}}}
\frac{\mbox{\rm exp}\left(\frac{2D}{|U|}\right)}{\sqrt{n(2-n)}} \,\,\,\,\, ,
\end{equation}
for (ii):

\begin{equation}\label{A14}
\frac{1}{\lambda^2(0)}\!=\!
\frac{16\pi e^2}{\hbar^2 c^2 a_{\mbox{\tiny $\bot$}}}
\frac{Z}{3}n(2-n)\frac{t^2}{|U|} \,\,\,\,\, ,
\end{equation}

\begin{equation}\label{A15}
\Delta_E\!=\!\frac{Dn(2-n)}{2}\left[\frac{|U|}{2D}+\frac{2D}{3|U|}-1\right]
\,\,\,\,\, ,
\end{equation}

\begin{equation}\label{A16}
\xi_{\mbox{\tiny G--L}}\!=\!\frac{a_{\mbox{\tiny $\|$}}t}{|U|} \,\,\,\,\, ,
\end{equation}
For $|U|/D\gg~1$ the results for $H_{\mbox{\tiny C}}$ ( and consequently also for
$\xi_{\mbox{\tiny G--L}}$ and
$\kappa$ ) are unrealistic as $E_0^{\mbox{\tiny NO}}$
in this limit largely overestimated by
MFA. As we have quoted in Sec. 2, one can correct these results by using
Eq.(\ref{EN-strongU}) instead of (\ref{A3}) in the calculation of
$H_{\mbox{\tiny C}}$.
%
%
%
%
%
%
\newpage
\noindent
Mailing address: Wojciech R. Czart, Institute of Physics, A. Mickiewicz
University, Umultowska 85, 61-614 Pozna\'{n}, Poland, e-mail:
czart@phys.amu.edu.pl. 
   
\newpage
\noindent
{\bf FIGURE CAPTIONS\\[1em]}
{\bf Fig. 1} 
Concentration dependence of $T_{\mbox{\tiny C}}$, the
gap parameter $\Delta$ and $1/\lambda^2$ at
$T$$=$$0$~K: $|U|/B$$=$$0.1$ ($B$$=$ bandwidth); {\bf a)} SC lattice, {\bf b)}
FCC lattice, {\bf c)} model rectangular density of states.
(here: $e^2/\hbar^2c^2a\!=\!1$, $a$--lattice constant, $B$--bandwidth).\\[1em]
{\bf Fig. 2} 
Concentration dependence of $1/\lambda^2$ at
$T$$=$$0$~K for different values of $U/B$ for SQ lattice.\\[1em] 
{\bf Fig. 3a} 
Reduced square value of thermodynamic critical field $H_{\mbox{\tiny C}}$ at
$T\!=\!0$ 
($h_{\mbox{\tiny C}}^2/B\!=\!H_{\mbox{\tiny C}}^2a^2/(8\pi B)$) 
as a function of
$|U|/B$ for $n\!=\!1$ for SQ lattice. HFA 
($\circ$), perturbation theory: $0^{th}$ order (obtained using
Eqs.(\ref{Hc},\ref{EN-strongU},\ref{MFAEss}) for
$J\!=\!0$) 
($\ast$), $2^{nd}$ order (obtained using
Eqs.(\ref{Hc},\ref{EN-strongU},\ref{MFAEss}) for $J\neq 0)$ (+).\\[1em]
{\bf Fig. 3b}
$h_{\mbox{\tiny C}}^2/B$ vs $n$ in $T\!=\!0$ plotted for increasing values of
$|U|/B$ in HFA for SQ lattice.
(numbers to the curves).
\\[1em]
{\bf Fig. 4} 
The G--L correlation length $\xi_{\mbox{\tiny G--L}}$ as a function of
$|U|/B$ for $n=1$ for SQ lattice. HFA (solid line),
perturbation theory: $0^{th}$ order
(obtained by putting $J\!=\!0$ in
Eqs.(\ref{xi},\ref{EN-strongU},\ref{MFAEss})) (dotted line),
$2^{nd}$ order (obtained using
Eqs.(\ref{xi},\ref{EN-strongU},\ref{MFAEss}) for $J\neq 0)$ (dashed line),
$T\!=\!0$ ($\xi_0\!=\!\frac{a_{\mbox{\tiny $\|$}}}{\sqrt{2}}$).\\[1em] 
{\bf Fig. 5}
$\xi_{\mbox{\tiny G--L}}(0)$ vs $n$ for SQ (full circles) and FCC
(triangles) lattices: 
$|U|/B\!=\!0.3$ .\\[1em]
{\bf Fig. 6} 
 Concentration dependence of $\kappa /\kappa_0$ 
 ($\kappa \!=\!\lambda/\xi_{\mbox{\tiny G--L}}$ at
$T\!=\!0$, $\kappa_0\!=\!\kappa (n\!=\!1)$)
 calculated for SQ lattice for several fixed
values of $|U|/B$ (numbers to the curves).
The corresponding plot obtained for a model rectangular DOS 
for arbitrary $|U|/B$ is shown by a dashed line.\\[1em]
{\bf Fig. 7} The boundary between local and nonlocal electromagnetic
behaviour in the $|U|/B$ - $n$ parameter space calculated from the equation
$\kappa \!=\!1/\sqrt{2}$ for $t\!=\!0.15$~eV, $a_{\mbox{\tiny $\|$}}\!=\!3.85$\AA ~and 
$a_{\mbox{\tiny $\bot$}}\!=\!7.7$\AA
 ~for rectangular DOS.\\[1em]
{\bf Fig. 8}
The values of $|U|/B$ for the attractive Hubbard model which provide the
Ginzburg ratio $\kappa \!=\!10^2$.
The plots as a function of $n$ for SQ lattice, $t\!=\!0.15$ eV, 
$a_{\mbox{\tiny $\|$}}\!=\!3.85$
\AA, $a_{\mbox{\tiny $\bot$}}\!=\!7.7$\AA ~(solid line) and for rectangular DOS
$t\!=\!0.15$eV, 
$a\!=\!7.7$\AA ~(dashed line).\\[1em]
{\bf Fig. 9} 
$\frac{T_{\mbox{\tiny C}}}{T_{\mbox{\tiny C}}^{\mbox{\tiny m}}}$ vs.
$\left(\frac{\lambda^{\mbox{\tiny m}}}{\lambda}\right)^2$ 
at $T$$=$$0$~K 
{\bf a)} for the SQ lattice, {\bf b)} SC lattice 
$\frac{|U|}{B}$$=$$2$~(---), 
$\frac{|U|}{B}$$=$$0.5$~($\cdot\cdot\cdot$),
$\frac{|U|}{B}$$=$$0.2$~(-~-)
and {\bf c)} for FCC lattice:
$\frac{|U|}{B}\!=\!5$ (-- --), $\frac{|U|}{B}\!=\!1$ (- -), 
$\frac{|U|}{B}\!=\!0.2$ (---).
 $\lambda^{\mbox{\tiny m}}$ corresponds to the
maximum 
transition temperature $T_{\mbox{\tiny C}}^{\mbox{\tiny m}}$. 
($\circ$)-Tl$_{\mbox{\scriptsize 2}}$Ba$_{\mbox{\scriptsize 2}}\!$
Ca$_{\mbox{\scriptsize 2}}$Cu$_{\mbox{\scriptsize 3}}\!$
O$_{\mbox{\scriptsize 10}}$, Tl$_{\mbox{\scriptsize 0.5}}\!$
Pb$_{\mbox{\scriptsize 0.5}}$Sr$_{\mbox{\scriptsize 2}}\!$
Ca$_{\mbox{\scriptsize 2}}$Cu$_{\mbox{\scriptsize 3}}\!$
O$_{\mbox{\scriptsize 9}}$, Bi$_{\mbox{\scriptsize 2-x}}\!$
Pb$_{\mbox{\scriptsize x}}$Sr$_{\mbox{\scriptsize 2}}\!$
Ca$_{\mbox{\scriptsize 2}}$Cu$_{\mbox{\scriptsize 3}}\!$
O$_{\mbox{\scriptsize 16}}$; ($\diamond$)-Y$_{\mbox{\scriptsize 1-x}}\!$
Pr$_{\mbox{\scriptsize x}}$Ba$_{\mbox{\scriptsize 2}}\!$
Cu$_{\mbox{\scriptsize 3}}$O$_{\mbox{\scriptsize 6.97}}$;
($\bigtriangleup$)-YBa$_{\mbox{2\scriptsize 2}}\!$
Cu$_{\mbox{\scriptsize 3}}$O$_{\mbox{\scriptsize x}}$;
($\bigtriangledown$)-La$_{\mbox{\scriptsize 2-x}}\!$
Sr$_{\mbox{\scriptsize x}}$CuO$_{\mbox{\scriptsize 4}}$;
($\star$)-Bi$_{\mbox{\scriptsize 2}}$Sr$_{\mbox{\scriptsize 2}}\!$
Ca$_{\mbox{\scriptsize 1-x}}\!$
Y$_{\mbox{\scriptsize x}}$Cu$_{\mbox{\scriptsize 2}}\!$
O$_{\mbox{\scriptsize 8+$\delta$}}$;
($\sqcap$)-LaMo$_{\mbox{\scriptsize 6}}$
Se$_{\mbox{\scriptsize2}}$, PbMo$_{\mbox{\scriptsize 6}}\!$
S$_{\mbox{\scriptsize 8}}$, SnMo$_{\mbox{\scriptsize 6}}\!$
S$_{\mbox{\scriptsize 4}}$Se$_{\mbox{\scriptsize 4}}$,
SnMo$_{\mbox{\scriptsize 6}}$S$_{\mbox{\scriptsize 7}}$Se,
SnMo$_{\mbox{\scriptsize 6}}$S$_{\mbox{\scriptsize 1}}\!$
Se$_{\mbox{\scriptsize 7}}$, LaMo$_{\mbox{\scriptsize 8}}\!$
S$_{\mbox{\scriptsize 8}}$, PbMo$_{\mbox{\scriptsize 6}}\!$
S$_{\mbox{\scriptsize 4}}$Se$_{\mbox{\scriptsize 4}}$;
~($\oplus$)-Tl$_{\mbox{\scriptsize 2}}\!$
Ba$_{\mbox{\scriptsize 2}}$ CuO$_{\mbox{\scriptsize 6+$\delta$}}$.
Experimental data taken from \cite{Niedermayer-93,Uemura-93}.
\\[1em]
{\bf Fig. 10}
$\frac{\lambda(n)}{\lambda(n\!=\!1)}$ at $T$$=$$0$~K for SQ lattice
($\frac{|U|}{B}$$=$$0.1$~($\ast$),
$\frac{|U|}{B}$$=$$0.5$~($\circ$)), compared with
experimental results (full squares) \cite{Locquet-95}
La$_{\mbox{\scriptsize 2-x}}$
Sr$_{\mbox{\scriptsize x}}$CuO$_{\mbox{\scriptsize 4}}$
films ( -$39$ nm) (squares),
$\lambda_{0}\!=\!\lambda(x\!\!=\!0.159)\!=\![6700\!\pm\! 500]$\AA.
The results for a model square density
of states almost coincide with the curve ($\circ$) for arbitrary $|U/B|$.

\end{document}